\documentclass[useAMS,usenatbib,usegraphicx]{mn2e}

\newcommand{\dg}{\mbox{$^\circ$}}
\newcommand{\hr}{\mbox{$^{\rm h}$}}
\newcommand{\am}{\mbox{$^{\prime}$}}
\newcommand{\as}{\mbox{$^{\prime\prime}$}}
\newcommand{\mass}{\mbox{M}}
\newcommand{\radius}{\mbox{R}}
\newcommand{\rjup}{\mbox{\radius$_{\rm Jup}$}}

\newcommand{\rsun}{\mbox{\radius$_{\sun}$}}
\newcommand{\msun}{\mbox{\mass$_{\sun}$}}
\newcommand{\mstar}{\mbox{\mass$_*$}}
\newcommand{\rstar}{\mbox{\radius$_*$}}

\newcommand{\rplanet}{\mbox{\radius$_p$}}
\newcommand{\rone}{\mbox{\radius$_1$}}
\newcommand{\rtwo}{\mbox{\radius$_2$}}
\newcommand{\mone}{\mbox{\mass$_1$}}
\newcommand{\mtwo}{\mbox{\mass$_2$}}
\newcommand{\msini}{\mbox{\mass$_p \sin i$}}

\newcommand{\epadu}{\mbox{$e^-$/ADU}}
\newcommand{\kmps}{\mbox{km s$^{-1}$}}

\newcommand{\ppl}{\mbox{P$_{\rm pl}$}}
\newcommand{\ptr}{\mbox{P$_{\rm tr}$}}
\newcommand{\pobs}{\mbox{P$_{\rm obs}$}}

\newcommand{\ndet}{\mbox{$N_{\rm det}$}}
\newcommand{\ndwarf}{\mbox{$N_{\rm dwarf}$}}
\newcommand{\nmin}{\mbox{$N_{\rm min}$}}

\newcommand{\sn}{\mbox{$S/N$}}
\newcommand{\snmin}{\mbox{$(S/N)_{\rm min}$}}
\newcommand{\trdep}{\mbox{$\Delta F$}}
\newcommand{\trdepmin}{\mbox{$\Delta F_{\rm min}$}}

\title[The UNSW Planet Search: Methods and First Results]
      {The UNSW Extrasolar Planet Search: Methods and First Results from a Field Centred on NGC~6633}

\author[M.~G.~Hidas et al.]{
  M.~G.~Hidas$^{1}$\thanks{E-mail: mgh@phys.unsw.edu.au},
  M.~C.~B.~Ashley$^{1}$, 
  J.~K.~Webb$^{1}$, 
  M.~Irwin$^{2}$, 
  A.~Phillips$^{1}$, 
  \newauthor 
  H.~Toyozumi$^{1}$,
  A.~Derekas$^{1,3}$,
  J.~L.~Christiansen$^{1}$,
  C.~Nutto$^1$,
  S.~Crothers$^{1}$
\\
$^1$~School of Physics, University of NSW, Sydney, NSW, 2052, Australia\\
$^2$~Institute of Astronomy, University of Cambridge, Madingley Road, Cambridge, CB3 0HA, United Kingdom\\
$^3$~School of Physics, University of Sydney, NSW, 2006, Australia }
\begin{document}

\date{29 March 2005}

\pagerange{\pageref{firstpage}--\pageref{lastpage}} \pubyear{2004}

\maketitle

\label{firstpage}

\begin{abstract}
We report on the current status of the University of New South Wales Extrasolar Planet Search project, giving details of the methods we use to obtain millimagnitude precision photometry using the 0.5~m Automated Patrol Telescope. We use a novel observing technique to optimally broaden the PSF and thus largely eliminate photometric noise due to intra-pixel sensitivity variations on the CCD. We have observed 8 crowded Galactic fields using this technique during 2003 and 2004. Our analysis of the first of these fields (centred on the open cluster NGC~6633) has yielded 49 variable stars and 4 shallow transit candidates. Follow-up observations of these candidates have identified them as eclipsing binary systems. We use a detailed simulation of our observations to estimate our sensitivity to short-period planets, and to select a new observing strategy to maximise the number of planets detected.
\end{abstract}



\section{Introduction}\label{sec:intro}

The detection of extrasolar giant planets via high-precision radial velocity measurements of dwarf stars in the Solar neighbourhood has been highly successful. Since the first detection \citep{MayorQueloz1995} over 100 planets have been found (e.g. \citealt{Butler2003}; \citealt{Tinney2003}; \citealt{Pepe2004}). About 20\% of these planets have periods of less than 10 days.\footnote{\it http://www.obspm.fr/encycl/catalog.html} These ``hot Jupiters'' have a significant probability ($\sim 10$\%) of having an orbital inclination that causes them to transit their host star. This makes possible an efficient photometric method of detecting such planets. In recent years a number of groups have begun photometric monitoring of large samples of stars, with the aim of detecting transiting hot Jupiters (see \citealt{Horne2003} for a list of projects). The first known transiting planet, HD~209458b, was found by the radial velocity method \citep{Charbonneau2000,Henry2000,Mazeh2000}. Only in the past year have transit searches finally begun to bear fruit, yielding five planets from the OGLE survey \citep{Udalski2002,Udalski2002b,Udalski2002a,Udalski2003,Konacki2003a,Bouchy2004a,Pont2004,Konacki2005a}, and the first confirmed detection by a small, wide-field telescope \citep{Alonso2004}.

Due to geometric and time-sampling considerations, the transit method has a lower probability of detecting a given planet around a star than the radial velocity (RV) technique. However, a search for transiting planets has considerable advantages.

\begin{itemize}

\item A transit search can be performed with a large-format CCD on a wide-field telescope, and can therefore monitor a large sample ($\sim 10^4$) of stars simultaneously. For planets with periods shorter than $\sim 10$ days, this outweighs the low transit probability and makes a transit search potentially more efficient than a radial velocity search, which necessarily targets fewer stars. Transit searches therefore have the capacity to build a larger sample of known close-in giant planets.

\item Measurements of RV variations in the host star yield most of the planet's orbital parameters, but only a minimum mass (\msini). For a transiting planet, the orbital inclination ($i$) can be estimated from the lightcurve (typically to within a few degrees), therefore the actual mass of the planet can be measured.

\item The radius of the planet can be measured from a fit to the transit lightcurve (and an estimate of the host star's radius), therefore the average density of the planet is known, providing an important constraint for models of its structure and composition.

\item Transit searches can be tailored to search for planets around stars in various environments, such as in open clusters \citep[e.g.][]{Mochejska2004,vonBraun2005}, globular clusters \citep[e.g.][]{Gilliland2000,Weldrake2005}, or in the Galactic disk. Radial velocity searches need to target relatively bright stars, and are therefore better suited to searching for planets in the Solar neighbourhood.

\item With ultra-high-precision photometry ($\la 0.1$~mmag), transiting Earth-like planets may be found \citep[e.g.][]{Deeg2002}. These will be key targets for planned space missions such as ESA's {\it Eddington} \citep{RoxburghFavata2003} and NASA's {\it Kepler} \citep{Borucki2003a}.

\item \citet*{RaymondQuinnLunine2004} have shown that terrestrial planets can form in the habitable zone (where liquid water can exist on the surface of a planet) of a star hosting a hot Jupiter. Assuming that the orbits in a planetary system are near co-planar, stars with known transiting hot Jupiters offer an excellent opportunity to detect (or rule out the existence of) terrestrial planets in the habitable zone.

\item While models suggest that planets may form and remain in stable orbits around binary stars \citep[e.g.][]{ArtymowiczLubow1994,HolmanWiegert1999}, no circum-binary planets are known to date.\footnote{With the possible exception of the system HD~202206, where the outer planet may have formed in a circum-binary disk around the primary and a brown dwarf companion \citep{Correia2004}.} Such planets could be found by targeting known eclipsing binary systems, which offer an increased probability of a transiting configuration (if the orbits are co-planar), as well as the possibility of detecting non-transiting planets via precise timing of binary eclipses \citep[and references therein]{DoyleDeeg2004}.

\item If the host star is sufficiently bright, the atmosphere of the planet can be studied using high-precision spectroscopy during a transit \citep[e.g.][]{Schneider1994a,SeagerSasselov2000,Brown2001b,WebbWormleaton2001}. To date, this has been done successfully for the planet orbiting HD~209458 \citep{Charbonneau2002,Vidal-Madjar2003,Vidal-Madjar2004}.

\item Infrared emission from a transiting planet may be detectable by the disappearance of spectral features when the planet is behind the host star \citep[e.g.][]{Richardson2003}.

\item Ultra-high-precision photometry and timing of planetary transits has the potential to detect moons and rings around giant planets \citep[e.g.][]{Brown2001a,DoyleDeeg2004,BarnesFortney2004}.

\item At a photometric precision of $10^{-5}$, a transiting gas giant's oblateness can have a detectable effect on the lightcurve, placing constraints on the planet's rotation rate \citep{SeagerHui2002}. Lensing of the host star's light by the planet's atmosphere may also be detectable \citep{HuiSeager2002}, providing further information about its structure.

\end{itemize}

Transit searches ideally require a wide-field telescope, in order to monitor a large sample of relatively bright stars. Transiting planets around stars bright enough for high-resolution spectroscopic follow-up are the most valuable detections.

The other key ingredient is the ability to obtain high-precision photometry. A typical hot Jupiter (radius 1~\rjup, orbiting a Sun-like star with a 3-day period), will cause a 1\% decrease in the host star's apparent brightness for $\sim3$~hours. Therefore a transit search needs to achieve a relative photometric precision of a few millimagnitudes over time-scales of at least a few hours, which is a challenge for CCD photometry with a wide-field, ground-based telescope. The present paper is primarily concerned with the methods we have used to reach this precision, and the first results of our search.

In section~\ref{sec:apt}, we describe the Automated Patrol Telescope and the camera. In section~\ref{sec:factors}, we identify the major factors limiting the photometric precision. The techniques we have used to overcome some of these limitations are discussed in sections \ref{sec:obs} and \ref{sec:analysis}. In section~\ref{sec:results} we present the data we have obtained to date, the quality of the resulting lightcurves, and the first results from our search. Estimates of our expected detection rate in the current data, as well as for our improved observing strategy, are given in section~\ref{sec:drate}.


\section{The Automated Patrol Telescope}\label{sec:apt}

The Automated Patrol Telescope (APT) at Siding Spring Observatory, Australia, is owned and operated by the University of New South Wales (UNSW). It is a 0.5~m telescope of Baker-Nunn design, with a 3-element correcting lens and an f/1 spherical primary mirror. It was originally used as a photographic satellite-tracking camera, before being donated to UNSW by the Smithsonian Institution, and converted for use with a CCD camera \citep{Carter1992}.

The camera was built by Wright Instruments, Enfield, U.K.. It uses an EEV CCD05-20 chip with $770 \times 1150$ (22.5~\micron) pixels to image a $2 \times 3$ degree field. However, the telescope has a potential field of view at least 5 degrees in diameter, and a new camera is being built in 2005 to take advantage of this.

The telescope is housed in a building with a roll-off roof. The operation of the roof, the telescope and the CCD camera are all controlled via a PC running Linux. An Internet connection to the PC gives a remote observer full control of the telescope. A local weather station and several web cameras allow the observer to monitor weather conditions and the operation of the telescope. The system also has the ability to close automatically in the event of rain, high humidity, or loss of connection to the Internet.


\section{Factors limiting photometric precision}\label{sec:factors}

We have used simulated APT images to study the factors contributing to the photometric noise. Here we summarise only the most important factors for bright stars.

\subsection{Poisson noise}

For unweighted aperture photometry, the variance in flux due to Poisson noise is $\sigma_{P,f}^{2} = (f + s) / g$, where $f$ and $s$ are the counts in the effective photometry aperture due to the star and sky respectively ($\sigma_{P,f}$, $f$, and $s$ all in ADU), and $g$ is the CCD gain ($g\approx 8$~\epadu\ for the APT). Expressed as an RMS variation in magnitude, provided $\sigma_{P,f} \ll f$, this becomes
\begin{equation}
\sigma_{P} \ \ = \ \ \frac{a}{f} \sqrt{\frac{f + s}{g}}
\end{equation}
where $a=2.5/\ln 10 \approx 1.086$, and the magnitude of the star is $m = z - 2.5 \log f$. The magnitude zero point, $z$, is the magnitude of a star which results in one ADU of detected flux at zero airmass. For a 2.5-minute APT image in $V$ band, $z\approx 22.5$. The precision we reach in practice is compared to this limit in Fig.~\ref{fig:rms}.

\subsection{Scintillation}

Scintillation sets the theoretical minimum noise level for the brightest (unsaturated) stars. The magnitude scatter due to scintillation is given by (\citealt{KjeldsenFransden1992}, equation~3)
\begin{equation}
\sigma_{\rm scint} = (0.09{\rm mag}) D^{-2/3} \chi^{3/2} \Delta t^{-1/2} e^{-h/8}
\end{equation}
where $D$ is the telescope diameter in centimetres, $\chi$ is the airmass, $\Delta t$ is the exposure time in seconds, $h$ is the altitude of the observatory in km. Using typical values for our observing program with the APT ($D=50$, $\chi=1.5$, $\Delta t=150$, $h=1.1$), the estimated scintillation limit is 0.9~mmag RMS per image.

\subsection{Undersampling and intra-pixel sensitivity variations}

The CCD currently used in the APT camera has 9.43~arcsec pixels. The image of a star in the focal plane has a FWHM equivalent to only $\sim 0.7$~pixel (Toyozumi \& Ashley, in preparation). 
It is far from a Gaussian profile, having a narrow, bright core, and broad, asymmetric wings (Fig.~\ref{fig:ipsf}, left panel). This ``instrumental point spread function'' (iPSF) is largely due to the telescope optics, and is therefore stable and largely independent of local seeing conditions. Its shape does, however, vary slightly with position on the CCD. When sampled by the CCD pixels, it becomes the ``effective PSF'' (ePSF), with a width of $\sim1.0$~pixel. For a focused star centred on a pixel, about 52\% of the light falls within that pixel. 

Because of this severe undersampling, the effect of CCD intra-pixel sensitivity variations (IPSV) becomes important (e.g. \citealt{Lauer1999}). Direct measurements by scanning the APT CCD with a small ($<0.2$~pixel) spot of light show that the point sensitivity varies by more than 30\% over a pixel in the $V$ band \citep{ToyozumiAshley2005:scan}. As a result, the total flux detected for a star varies with its precise position within a pixel (Fig.~\ref{fig:ipsv}, left panel). Combined with inevitable small shifts due to telescope jitter and differential refraction, this constitutes a significant source of photometric noise. Since IPSV is largely caused by the surface structure of the CCD, its amplitude decreases with increasing wavelength. For focused images, the detected flux varies by 4\% in $V$ band, and 2.5\% in $I$ band. Due to the uniform structure of the CCD, the sensitivity variation is, to a good approximation, identical in all pixels.

Conventionally, PSF fitting is the technique of choice for general purpose precision photometry. However, with undersampling, IPSV, and variation of the PSF over the field of view, photon-noise limited PSF fitting is extremely difficult to achieve. In principle these effects can be accounted for, and practical methods for doing so have been described by \citet{Lauer1999} and \citet{AndersonKing2000}. 
We have applied the algorithm of Anderson \& King to APT images. While the the result was a significant improvement over naive aperture photometry, a precision below $\sim 1$\% could not be reached for most stars \citep{Hidas2003a,Hidas2004}. The severe undersampling of APT images is the likely cause of this limit. Most of the objects to be measured are blends of several unresolved stars. Thus the true number of PSFs contributing to the observed pixel flux distriubtion is often significantly higher than what the model is fitting (or is capable of deriving from the image itself).

\subsection{Flatfielding}

We have made sure that flatfielding errors do not make any significant contribution to the photometric noise for bright stars. Any changes in the pixel-to-pixel sensitivity variations over time-scales of a few months are at a level below 0.1\%. We combine a sufficient number of twilight flatfield images so that the Poisson noise in the combined master flat is also well below 0.1\% RMS per pixel.

Note that as twilight skies have generally bluer optical colours than
dark skies, while most target stars generally have redder colours, the
pixel response as a function of wavelength also needs to be taken into
account.  However, this is not a problem if only the overall
sensitivity is dependent on wavelength, and the relative
pixel-to-pixel differences are constant.  This seems to be the case
for most CCDs.

\subsection{Non-linearity of CCD response}

The relationship between the flux incident on the CCD and the resulting pixel values can be significantly non-linear, largely due to the CCD readout amplifier. This can be another limiting factor for precision differential photometry from undersampled images, where the distribution of pixel values in the photometry aperture can vary considerably from image to image. 

\begin{figure}
  \centering
  \includegraphics[width=63mm]{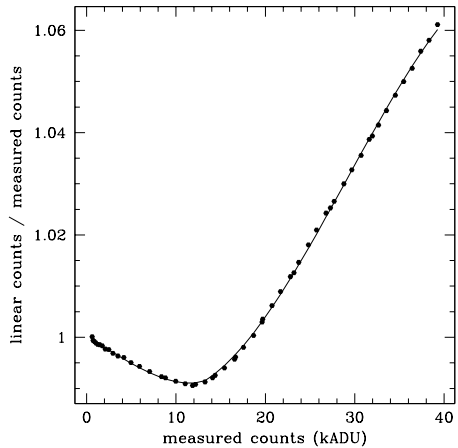}
  \caption{Deviation from linearity of the CCD response, measured from a set of dome flat images taken at a range of exposure times. The fitted curve (two third-order polynomials joined at $\sim 13000$~ADU) is used to correct images for this effect. }
  \label{fig:nonlin}
\end{figure}

Fig.~\ref{fig:nonlin} shows our measurement of the response of the APT CCD to a range of incident flux levels (in flatfield images). The signal in bright pixels deviates from linearity by up to 6\%, which is large compared to many CCD cameras. We have fitted two third-order polynomials (above and below 13000~ADU) to these data, and use this fit to correct every image before further processing. As the residuals from the fit are less than 0.05\%, this correction reduces the effect of non-linearity on the photometric noise to below 0.5~mmag.

\subsection{Crowding}\label{ssec:crowding}

In a crowded field, photometry apertures placed around stars frequently have relatively bright pixels near their boundaries. These are from blended or nearby stars within a distance of $\sim 30$\as\ (the radius of the aperture we use). Small shifts in position ($\sim 1$~pixel), due to telescope jitter and image rotation, cause variations in both the brightness of such pixels, and the extent of their overlap with the aperture. The amplitude of this additional noise will depend on each individual star's immediate neighbourhood in the field. In the worst case, the presence of a relatively bright neighbour can add 10--20~mmag (RMS) of noise. This can occur for all stars, but will occur more commonly for the fainter stars. 

Furthermore, the presence of faint, unresolved background stars makes the effective sky level more noisy and variable over short spatial scales. Thus, while crowding does not place a strict limit on the precision, on average it increases the level of noise in the lightcurves.

In order to minimise the additional photometric noise due to crowding, we have experimented with allowing each object to have a different sized photometry aperture. The aperture radius is chosen to minimise the flux in the brightest pixel near the boundary in a sample image, while applying a weight towards smaller apertures. Once a radius is chosen for an object, it is kept fixed for all measurements. For most stars fainter than $V=14.5$ this reduces the RMS scatter by at least 20\%. To date we have found no clear improvement for the brighter stars, though future refinement of this idea may lead to significant gains.

Besides contributing to the photometric noise, crowding has another important effect on our ability to detect variations in a star's magnitude (including transits). The additional light from blended stars ``dilutes'' the amplitude of any variations, reducing the signal to noise ratio of planetary transits, and making the eclipses of binary stars appear more like transits (see section~\ref{ssec:falsies}).

\subsection{Atmospheric conditions}

Relative photometry is generally not affected by temporal variations in overall sky transparency over the course of a night. However, the transparency can also vary spatially, for example due to undetected thin clouds. For a wide-field instrument like the APT, this can result in significant variations in magnitude differences across the field, and thus additional noise in the relative photometry. When such ``non-photometric'' conditions occur, they can degrade the precision for bright stars ($9 \la V \la 13$) to 5--10~mmag, making the data obtained only marginally useful.

\subsection{Systematic errors}\label{ssec:syserr}

Systematic variations in apparent magnitude common to all stars are removed by calibrating the photometry using a large sample of stars (see section~\ref{ssec:calib}). However, in some of the resulting lightcurves, systematic trends do remain. These are usually in the form of a linear change in apparent magnitude with time, amounting to a $\sim10$~mmag decrease or increase over the course of a night. We note that similar trends have been observed in lightcurves from other photometric surveys, such as HATNet \citep{Bakos2004}, MACHO \citep{DrakeCook2004}, and OGLE \citep{Udalski2003}, though the latter authors also claim that these can be removed using a method developed by \citet{KruszewskiSemeniuk2003}.

In our own data, these systematics may be largely due to the inability of our sky fitting routine to accurately follow changes in the effective background signal over small spatial scales. This can occur in regions of the field where the density of faint, unresolved background stars is large (e.g. near the centre of an open cluster). Combined with the time-dependent slope in the intrinsic sky brightness, this can lead to the kind of trends we observe. 
We are working on an improved sky fitting algorithm in order to minimise these effects.

Small changes in the camera focus also lead to systematic errors in the relative photometry, as they alter the contribution of light to each star's photometry aperture from blended or nearby stars. Since the magnitude of this change varies depending on each star's immediate neighbourhood, it cannot be simply corrected for. While the focus is stable over the course of a typical observing run, it does need to be reset occasionally, in particular after a filter change. This effect can cause magnitude offsets from night to night in the lightcurves of some stars (those whose fractional contribution from neighbours is significantly different from the reference stars used for calibration).


\section{Observing Strategy}\label{sec:obs}

\subsection{The raster-scan technique}\label{ssec:raster}

Tests with severely defocused images (FWHM $\approx 10$~pixels) have shown that a relative photometric precision of at least $\sim 3$~mmag was possible with the APT, for bright ($V \approx 7$) stars in sparse fields. Of course, such images are not appropriate for observing a large number of fainter stars in crowded fields. 

\begin{figure}
  \includegraphics[width=84mm]{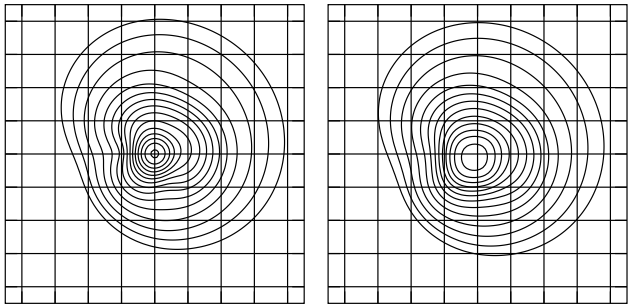}
  \caption{ The instrumental PSF (iPSF) of the APT near the centre of the field. Each PSF has been normalised to have a total flux of 1. The contour levels are identical in both panels, and have uniform logarithmic spacing. The grid spacing corresponds to 1 CCD pixel. {\em Left panel:} The normal, focused iPSF, determined by fitting a multi-Gaussian model to a set of 150-sec, $V$-band  images of a bright star (Toyozumi \& Ashley, in preparation). 
The FWHM is $\sim 0.7$~pixel and the peak intensity $\sim 1.1$. {\em Right panel:} The broadened iPSF obtained by the raster-scan technique. This model was calculated by numerically convolving the normal iPSF with the raster-scan pattern followed by the telescope (Ashley et al., in preparation). It has a FWHM of $\sim 1.5$~pixels, and a peak intensity $\sim 0.5$. }
  \label{fig:ipsf}
\end{figure}

\begin{figure}
  \includegraphics[width=84mm]{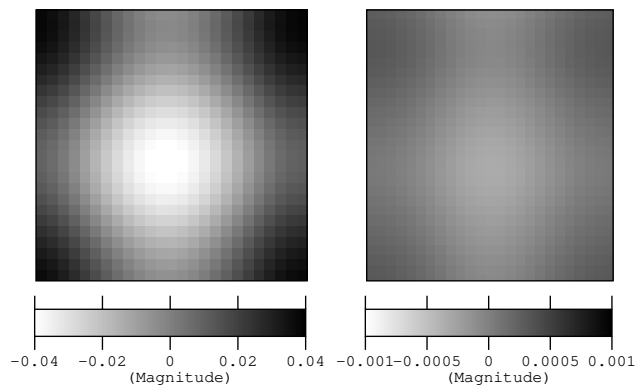}
  \caption{ The effect of intra-pixel sensitivity variations (IPSV) on the photometry from undersampled images taken by the APT. Each image represents one CCD pixel, and shows the relative magnitude measured for a star, as a function of its position within the pixel. An accurate model of the instrumental PSF, obtained from real images (Toyozumi \& Ashley, in preparation), 
was convolved with the pixel sensitivity, as measured by scanning the CCD with a spot of light \citep{ToyozumiAshley2005:scan}. The two panels correspond to the two iPSF models presented in Fig.~\ref{fig:ipsf}. {\em Left panel:} the result for normal, focused APT images in the $V$ band. {\em Right panel:} the reduced effect when the PSF is broadened by the raster-scan technique (section~\ref {ssec:raster}). Note the difference in scale. }
  \label{fig:ipsv}
\end{figure}

A new observing technique has allowed us to achieve the same precision without degrading image quality (Ashley et al., in preparation). During each exposure, the telescope is moved in a precise raster-like pattern covering the area of a single CCD pixel. This broadens the effective PSF only slightly (FWHM $\approx 1.5$~pixels, see right panel of Fig.~\ref{fig:ipsf}). However, because each point on the instrumental PSF scans over a range of sub-pixel positions during the exposure, the sensitivity variations within the pixel are effectively averaged out (Fig.~\ref{fig:ipsv}). Residual variations in the detected flux due to IPSV are limited to $\la 1$~mmag. 

A slightly different technique, based on the same idea of broadening the PSF by moving the telescope during exposure is used by the Hungarian Automated Telescope \citep{Bakos2004}.

\subsection{Nightly routine}

Most of the observations are conducted remotely via the Internet. The observer's main task is to initiate the observing process, and monitor weather conditions and the quality of the data obtained. Each of the tasks of starting up the telescope system, opening the roof, taking twilight flats, and the observations themselves are executed via simple shell scripts on the telescope-control PC. 

Twilight flatfield images are taken on most clear evenings when the Moon is down, and on some mornings. $\sim 15$ images with sufficiently high background flux can be taken per twilight. Other calibration images are not necessary for the APT camera. The bias level shows no significant structure over the CCD, and can be accurately measured from the overscan region. Dark current is negligible ($\sim 0.6$~e$^-$ per pixel per second).

During the night, as each image is read out, an astrometry program locates Tycho-2 stars \citep{Hog2000} in the image, and calculates the (linear) transformation from CCD pixel coordinates to celestial coordinates. The coefficients of the transformation are written to the FITS header as a World Coordinate System (WCS). Typically, $\sim 200$ Tycho-2 stars are matched. In this way, telescope pointing is kept accurate to within $\sim 0.2$ of a pixel. This helps reduce the effects of any flatfielding errors, as stars near the centre of the field are always imaged by the same pixels. As differential refraction by the atmosphere causes a gradual rotation ($\sim 4$~arcmin in 8~hours) and change in image scale ($<0.1$\%), stars near the corners of the field will shift by $\sim 1$~pixel.

The astrometry program also performs a quick measurement of the magnitudes of the Tycho stars and compares them to the catalogue values. This allows the remote observer to monitor sky transparency and detect the presence of clouds.

For most of our 2003--2004 observations, the telescope continuously cycled between four adjacent fields, taking 150-sec exposures, as long as the fields were observable. Including CCD readout and telescope slew time, each image took 3.5~minutes. On a full clear night $\sim 30$ images per field were obtained. Due to physical limitations of the mount, the telescope can follow a field for a maximum of $\sim 8$ hours.


\section{Analysis Strategy}\label{sec:analysis}

\subsection{Data reduction}

Optimal aperture photometry can be competitive with PSF fitting for
moderately crowded fields \citep[e.g.][]{Naylor1998}. For brighter
stars, variance-weighted PSF fitting is equivalent to standard
aperture photometry \citep[e.g.][]{Irwin1997}.  This and the constancy
of the PSF for the APT have enabled us to develop a simple, robust
and extremely accurate aperture photometry package for processing APT
images.  Image crowding, usually the bugbear of aperture photometry,
is dealt with in a novel, yet simple manner.

Briefly, each APT frame is processed in the standard manner up to and
including generation of object catalogues using the development toolkit
of the INT WFS pipeline software \citep{IrwinLewis2001}. Particular care
is taken over flatfielding, since we need to ensure the gross
pixel-to-pixel sensitivity variations are mapped to better than the
0.1\% level. We combine all the twilight flats (of order 100) obtained
during a 3--4 month season. Each twilight image has $\sim20000$
counts (the gain is $\sim8$~\epadu).

Non-linearity of pixel counts is dealt with by first measuring the
global non-linearity from a varying time series of, essentially,
flatfield frames under constant illumination and then applying the
derived constant look-up-table correction to all pixels before
processing.

The final stages of the INT WFS pipeline involves generating object
catalogues for each frame.  Among other things the object catalogues are
used to derive accurate 6-parameter linear frame-to-frame transforms
with respect to a chosen master reference frame. To date, a single,
low-airmass image from a dark, photometric night has been chosen as
the master frame for each field. In the future a separate master image
will be generated by stacking a dozen or so consecutive images. This
will remove spurious objects from the master catalogue (such as cosmic
rays), and improve the astrometric precision.

The transformations are of the general linear form
\begin{equation}
x' = ax + by + c\;;  \ \ \ \ \   y' = dy + ex + f 
\end{equation}
where $a$ and $d$ define the relative $x$ and $y$ scales, $b$ and $e$
the rotation and/or shear, and $c$ and $f$ the global shifts. Note that we
could perform these transformations using the WCS data stored in each
image header. However, this would not make optimal use of the information
in each frame (the WCS transformations are calculated using $\sim 200$
Tycho stars, while the full object catalogues typically contain 10000
objects to be matched to the master catalogue), and would limit the
accuracy to $\sim0.1$~pixel.

Although we attempt to place the exposures of each target on
comparable field centres within each night and from night to night by
performing astrometry on each image as it is read out, small
differential adjustments of order 1 pixel are still required.  Even
with apparently perfectly aligned images, small temperature changes
throughout nights, differential refraction variations as a function of
airmass and differential field rotation all contribute significant
pixel level distortions.  However, the approximate global alignment
does ensure that non-linear terms are negligible at this pixel scale.

The reference frame catalogue also defines the master coordinate list to
be used for the later analysis, and this is transformed (to within
0.01 of a pixel) onto each frame to be processed.  Without this stage
relative accuracies of $\sim 0.1$ of a pixel are still possible, from
the individual image catalogues.  However, with these undersampled
images, coordinate accuracies of 0.1~pixel are not sufficient to give
mmag photometry for the brighter objects. This is particularly
important in crowded fields, where bright pixels on the boundary of
the photometry aperture are frequent (see
\ref{ssec:crowding}). Variations in the contribution to the aperture
sum from these pixels due to errors in aperture centring need to be
minimised. Tests have shown that errors of $\sim0.1$~pixel in the
aperture positions increase the photometric noise by 50\% for the
average bright ($V \la 14$) star in a crowded field (compared to the
near-photon-limited results we otherwise achieve).

The final ingredient is a very stable and robust 2D background
estimator, which is used to estimate and remove the global background
in a highly repeatable manner in each frame.  With this kit of parts
each dataset can be analysed in (almost) exactly the same way, using
the same apertures accurately co-located on each frame.  Each aperture sum
includes sub-pixel partitioning of flux and the whole package runs
completely automatically.

The main advantage of this approach is that systematics from crowding are
the same in each frame because of the PSF stability and the differential 
coordinate precision, and even quite crowded fields can be successfully 
analysed since essentially identical image confusion arises in all cases.
With this approach we are achieving differential
photometric precision of better than 2~mmag for bright objects and are close 
to the theoretical photon noise limit at all magnitudes (Fig.~\ref{fig:rms}). 

\subsection{Photometric calibration}\label{ssec:calib}

Global image-to-image magnitude variations are removed by ensemble relative photometry. An approximate magnitude zero-point (good to $\sim0.1$ mag) is set by comparison with the magnitudes of 100--200 stars from the Tycho-2 catalogue. Stars in the range $9 < V < 11$ are then used for the calibration. There are typically $\sim 1000$ of these in a dense Galactic field. Magnitude residuals ($\Delta m$) from the nightly mean value for each reference star are calculated, and for each image, fitted by a function of the form
\begin{equation}
\Delta m = a + bx + cy + dxy + ey^2
\end{equation}
where $x, y$ are the star's pixel coordinates on the CCD, and $a$--$e$ are constants for an image. A second order term in $y$ was necessary because the CCD is larger along the $y$-axis, which corresponds to the right-ascension direction (along which the airmass varies more). Adding an $x^2$ term to the fit does not significantly improve the photometric precision. Therefore, to make the fit more robust, this term is not included. As this function provides a good approximation to the variation in airmass across the field, a separate airmass term is not necessary.

This fit is then subtracted from each star's magnitude. Mean magnitudes and residuals and recalculated. Based on the RMS residuals of each star over the night's data, variable and badly measured stars are removed from the reference list, and the whole process is repeated. In most cases, the coefficients of the fit are effectively zero after 4--7 iterations.
Any images for which the median RMS of stars with $9 < V < 12.5$ remains higher than 15~mmag after calibration are rejected, and the calibration is repeated without them.

The above process is applied to each night's data separately. Any remaining night-to-night magnitude zero-point offsets are removed at the time the final lightcurves are compiled.

\subsection{The pipeline}

The entire data reduction process, from raw images to calibrated magnitudes, can be run automatically using a Perl script. The only interactive part of the procedure is the selection and combining of a suitable set of twilight flats to create a master flatfield image. This needs to be done once for each 3-4 month season of data. Running on a Sun Enterprise 4500 
workstation (with 8 CPUs and 4~Gb of memory), the pipeline will then process a full night's data in less than 2 hours, and several of these processes can be run in parallel.

From each dataset, two sets of magnitudes are generated, measured with photometry apertures 2 and 3 pixels in radius. The larger aperture gives a slightly lower average RMS for most bright stars, but is more susceptible to contamination of lightcurves by close neighbours. When a small-amplitude signal, such as a transit, is found in a lightcurve, a comparison of the signal amplitude measured by the two apertures provides a quick way of identifying such contamination.

\subsection{Candidate selection}\label{ssec:cansel}

Initially, we visually inspect the lightcurves for the brightest 2--3000 stars in each field (down to $V \approx 13$). This procedure has the advantage of being less susceptible to spurious detections due to variable stars and systematic errors than detection software.

We also use the transit detection algorithm described in \citet{AigrainIrwin2004}. Lightcurves are first searched for individual, box-shaped transit events of a given trial duration, and the signal-to-noise ratio (\sn) of a transit at each trial epoch is stored. These \sn\ values are then used to perform a period search. This is repeated for a number of trial transit durations, and the combination of period, duration and starting epoch resulting in the highest total \sn\ (over all observed transits) is selected for each lightcurve. A number of selection criteria are applied, based on the individual transit and total \sn, as well as the distribution of \sn\ values (over all trial parameters) for each lightcurve.

To date, the effectiveness of this detection software has been limited by the presence of systematic trends in our lightcurves (see section~\ref{ssec:syserr}), which lead to frequent spurious detections. \citet*{KovacsBakosNoyes2004} have recently published details of a trend filtering algorithm which substantially improves the quality of lightcurves obtained by the HATNet project \citep{Bakos2004}. Our initial experiments applying this algorithm to APT data had promising results, though our implementation is yet to be refined. At present it does greatly reduce the amplitude of systematics while preserving most transit signals, but also increases the white noise somewhat. We have not yet fully characterised the sensitivity of our detection software, but preliminary tests have shown that in the absence of systematics, 10~mmag transits are detected in a high percentage of cases (provided multiple transits are present and the total \sn\ is sufficiently high).

\subsection{Eliminating false positives}\label{ssec:falsies}

\citet{Brown2003} has estimated that up to $\sim$90\% of planet candidates selected by a wide-field transit search are binary stars. Grazing eclipses of binary stars, and eclipsing binaries with their light diluted by a third blended star can mimic the depth, and (to some extent) the shape of a planet transit. It is essential to eliminate these false positives at the earliest possible stage of the follow-up process. Our follow-up strategy is outlined below. At each stage we focus on the most promising planet candidates. Depending on the available observing time we also follow up likely binary systems.
\begin{enumerate}

\item Inspection of a Digitised Sky Survey\footnote{The Digitised Sky Survey was produced at the Space Telescope Science Institute under U.S. Government grant NAG W-2166, based on photographic data obtained using the Oschin Schmidt and UK Schmidt telescopes.}
 image, to reveal how much blending is present in the APT image (our photometry apertures are effectively $\sim$1~arcmin in diameter, which is typical for wide-field transit search projects).

\item Estimation of the physical parameters of the system, in particular the radius of the transiting object, from the lightcurve. The lightcurve is parametrised by the period ($P$), transit depth (\trdep), and the durations of the complete transit and the ``flat bottom'' part ($t_T$ and $t_F$). From these, the primary star's density can be calculated \citep{SeagerMallen-Ornelas2003}. Assuming a stellar mass-radius relation then yields the primary's radius (\rone), and the radius of the transiting object $\rtwo = \rone \sqrt{\trdep}$. Assuming the transit is in front of the brightest star in the APT aperture and adjusting the transit depth accordingly gives a lower limit on the size of the transiting object.

\item Photometric monitoring of the target area at higher spatial resolution, during a predicted transit, to identify the host of the transiting object, and measure the true (or at least less diluted) transit depth. This is done in multiple colours so that the colour signature of a grazing or blended eclipsing binary can be detected.

\item Medium-resolution spectroscopic follow-up to identify the host star's spectral class, allowing a more precise estimate of its radius, and therefore the radius of the planet. The presence of double absorption (or emission) lines, features from different spectral types, and in particular large radial velocity variations in phase with the transit signal, would indicate a binary or blend.

\item For the best candidates, high-precision radial velocity measurements (using echelle spectroscopy) will allow the measurement of the transiting object's mass and confirm or rule out its planetary nature.

\end{enumerate}


\section{Results}\label{sec:results}

\subsection{Fields observed}\label{ssec:fields}

We have targeted two groups of four adjacent fields, one group in Centaurus, the other in Ophiuchus. Both groups are within 10 degrees of the Galactic plane. The field centres are shown in Table~\ref{tab:fields}. We have centred one field in each group on an open cluster (NGC~3532 and NGC~6633). However, as each cluster contains only a few hundred relatively bright stars \citep[e.g.][]{Koelbloed1959,Jeffries1997}, and many of these are blended or saturated, only a small fraction of the stars we measure are cluster members.

\begin{table}
  \caption{Fields targeted in the UNSW planet search.}
  \label{tab:fields}
  \begin{tabular}{@{}lcccccc}
    \hline
    Field name & R.A.& Dec.  &  $b$  &  $N_*(V<14)$  &  $N_{\rm nights}$  \\
    \hline
    NGC~3532 & 11\hr 07\am &$-$58\dg 42\am & 1\dg & 7000 & 61 \\
    G1       & 11\hr 31\am &$-$58\dg 42\am & 3\dg & 7000 & 65 \\
    G2       & 11\hr 31\am &$-$56\dg 30\am & 5\dg & 7000 & 60 \\
    G3       & 11\hr 07\am &$-$56\dg 30\am & 4\dg & 7000 & 59 \\
    NGC~6633 & 18\hr 28\am &$+$6\dg 35\am & 8\dg & 5000 & 79 \\
    H1       & 18\hr 28\am &$+$4\dg 30\am & 7\dg & 4000 & 47 \\
    H4       & 18\hr 41\am &$+$4\dg 30\am & 4\dg & 3000 & 30 \\
    H5       & 18\hr 41\am &$+$6\dg 35\am & 5\dg & 5000 & 30 \\
    \hline
  \end{tabular}
\end{table}

We began observing most of these fields in 2003, and we now have two full seasons' worth of data for each of them. The first observations using the method described in section~\ref{sec:obs} were taken in August 2002, targeting the open cluster NGC~6633 as part of a multi-site campaign \citep{Martin2004}. A total of 690 images were taken through a Johnson $V$ filter on eight consecutive nights (7--14 August). In 2003 we added the three adjacent fields, and obtained another 180 images of NGC~6633 in $V$ band (June), and 380 in $R$ band (July--August). During the 2004 season we continued to observe the field in $V$ band (March--July, 1800 images), including some nights where we again observed this field only, to obtain higher-cadence measurements. About 1/3 of the images taken in 2004 (typically those taken in the first part of the night) were affected by a technical problem which has rendered them largely unusable. Finally in August 2004 we began observing in $I$ band, and took 230 images of NGC~6633. For these latter observations, the exposure time was reduced to 60 seconds, as the CCD is more sensitive in this band.

Due to its Northern declination, the NGC~6633 field can be observed for a maximum of 7 hours at relatively low airmass. From experience, useful images can be taken up to an airmass of $\sim 2.2$. The minimum airmass reached by this field is 1.3.

With a threshold of $4 \sigma$ above the background, the pipeline software typically detects $\sim 9000$ objects in the field, the faintest of these having $V \approx 15.5$. In each image, the positions of $\sim 7000$ of these objects are matched to the master catalogue to obtain the coordinate transformations into that image, with mean residual errors of $\sim 0.05$~pixel.

\begin{figure}
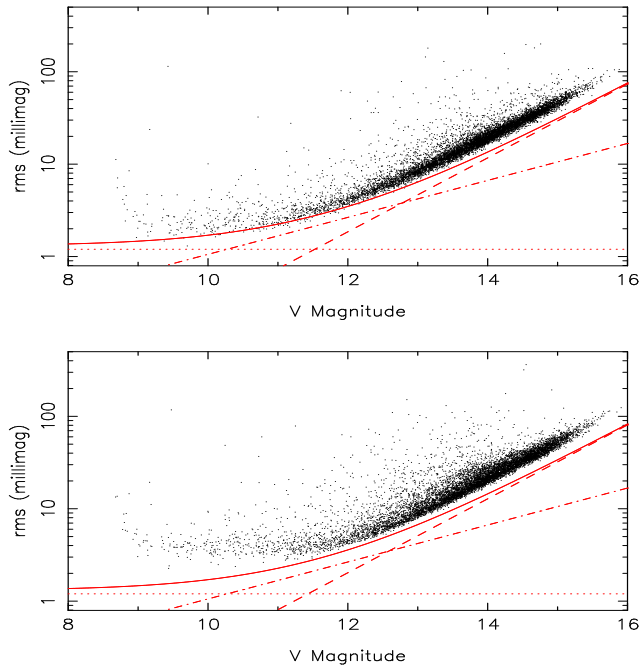

  \includegraphics[width=42mm,height=84mm,angle=-90]{rms1.ps}
  \\
  \\
  \includegraphics[width=42mm,height=84mm,angle=-90]{rms5.ps}
  \caption{ Photometric precision from one night (top panel), and 5 nights (bottom panel) of APT observations of the field centred on the open cluster NGC~6633 (August 2002). The solid line is the theoretical estimate, including Poisson noise in the star flux (dot dashed) and sky flux (dashed), and a magnitude-independent term (dotted). The latter term (1.2~mmag) represents scintillation, as well as any residual errors due to flatfielding, non-linearity, and IPSV. Stars brighter than $V \approx 9$ are saturated in a 150-second APT image. }
  \label{fig:rms}
\end{figure}

The relative photometric precision achieved on a single clear night, and over a week of observations, is shown in Fig.~\ref{fig:rms}. For stars in the range $9 \la V \la 11$, we routinely achieve 2~mmag precision. For this field, the nightly RMS is under 10~mmag for 2300 stars.

\subsection{Transit candidates}

\begin{figure}
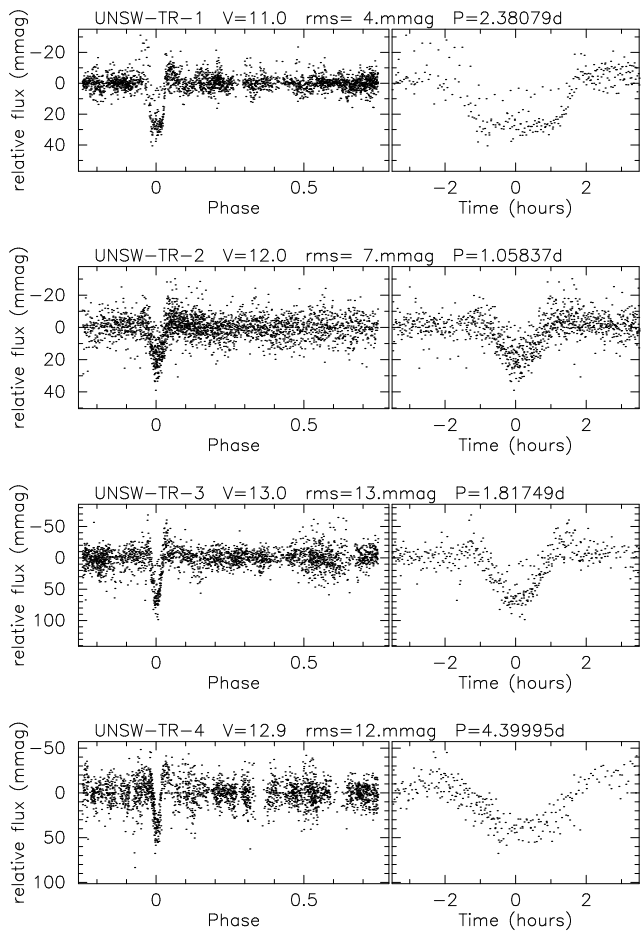

  \includegraphics[height=84mm,angle=-90]{08275.ps}
  \\
  \\
  \includegraphics[height=84mm,angle=-90]{10467.ps}
  \\
  \\
  \includegraphics[height=84mm,angle=-90]{11140.ps}
  \\
  \\
  \includegraphics[height=84mm,angle=-90]{12053.ps}
  \caption{ Phase-folded lightcurves of the 4 transit candidates found in the NGC~6633 field. The lightcurves include all $V$ band data of reasonable photometric quality (nightly rms $\la$ 10~mmag and at least 10 points per night) we obtained in 2002--2004. Note that UNSW-TR-2 is folded at its originally estimated period. Follow-up observations have shown that this is a binary system with two nearly identical eclipses, and the true period is twice this value. }
  \label{fig:candidlc}
\end{figure}

The lightcurves for the brightest $\sim 2000$ stars (down to $V \approx 13$) from the NGC~6633 field were searched for transit-like events by visual inspection. Through this process we identified 4 lightcurves with transits shallow enough ($<100$~mmag) to be considered as potential planet candidates. Their lightcurves are shown in Fig.~\ref{fig:candidlc}. 

\begin{table*}
\begin{minipage}{170mm}
  \caption{Transit candidates in the NGC~6633 field: coordinates and parameters estimated from APT data. $T_c$ is the time at the centre of the first transit we detected (JD-2450000).  Coordinates are from the WCS solution in the APT images, and represent the centroid (to within $\sim 1$\as) of the object, which is generally a blend. Similarly, the given $V$ magnitudes, with errors of $\sim 0.1$~mag, are for the blend as measured by the APT.  Errors in the other measured parameters are $5 \times 10^{-5}$~d for the period, 0.01~d for $T_c$, 5~mmag for the transit depth ($\Delta F$), and $\sim 0.5$~hr for $t_T$ and $t_F$ (the duration of the complete transit, and of the ``flat bottom'' part, respectively). Estimates for the radius of the star (\rstar) and the planet (\rplanet) were derived using the equations of \citet{SeagerMallen-Ornelas2003}. UNSW-TR-1 is a member of the cluster (NGC~6633~141).}
  \label{tab:candidates}
  \centering
  \begin{tabular}{lcccccccccc}
    \hline
         ID   &    R.A.    &    Dec.   & $V$  & $P$     & $T_c$   & \trdep& $t_T$ & $t_F$ & \rone   & \rtwo  \\
              &   (J2000)  &  (J2000)  &      & (days)  & (days)  &  (mmag)   & (hr)  & (hr)  & (\rsun) & (\rjup)  \\
    \hline				    		 	 		      	      				  
    UNSW-TR-1 & 18 29 04.4 & +06 26 54 & 11.0 & 2.38079 & 2496.93 &    30     & 3.4   & 2.0   & 1.4     & 2.9      \\
    UNSW-TR-2 & 18 30 51.9 & +07 09 20 & 12.0 & 1.05837 & 2462.99 &    25     & 1.9   & 0.2   & 1.5     & 2.4      \\
    UNSW-TR-3 & 18 31 00.7 & +07 08 25 & 13.0 & 1.81749 & 2499.07 &    70     & 2.0   & 0.5   & 0.8     & 2.0      \\
    UNSW-TR-4 & 18 32 22.3 & +06 37 13 & 12.9 & 4.39995 & 2465.13 &    50     & 3.8   & 0.7   & 1.2     & 3.8      \\
    \hline
  \end{tabular}
\end{minipage}
\end{table*}

\begin{table}
  \caption{Transit candidates in the NGC~6633 field: parameters obtained from follow-up spectroscopy. The error on the primary mass (\mone) was assumed to be 10\%. The velocity amplitude of TR-2 was estimated from the separation of double lines in the single spectrum we obtained. As it is a binary system with nearly identical components the orbital period is twice the value measured from the APT lightcurves. We assumed the two stars have equal mass. The orbit of TR-4 appears to be eccentric, but we have insufficient data for a rigorous Keplerian fit. Assuming a circular orbit gives the result quoted here.}
  \label{tab:followup}
  \centering
  \begin{tabular}{lcccc}
    \hline
    Star ID    & Sp Type & \mone    & $K$        &\mtwo \\
               &         & (\msun)  & (\kmps)    &(\msun) \\
    \hline	     	 	 
    UNSW-TR-1  & A2V     &   2.5    & $27\pm3$   & $0.34\pm0.05$ \\
    UNSW-TR-2  & K7V     & $=\mtwo$ & $87\pm3$   & $0.58\pm0.05$ \\
    UNSW-TR-3  & F7V     &   1.2    & $53\pm10$  & $0.4\pm0.1$   \\
    UNSW-TR-4  & F5V     &   1.3    & $\sim20$   & $\sim0.2$     \\
    \hline
  \end{tabular}
\end{table}

Table~\ref{tab:candidates} summarises the properties of the candidates, including an estimate of the transiting object's minimum radius (see item (ii) under section~\ref{ssec:falsies}). With relatively large radii, our candidates seemed unlikely to be planets. The largest extrasolar planet known to date (HD~209458b) has a radius of 1.35~\rjup\ \citep{Brown2001a}. However, as \citet{DrakeCook2004} point out, there are currently too few transiting objects with known mass to place a firm upper limit on the radii of extrasolar planets and brown dwarfs. Though not our primary targets, brown dwarfs and low-mass stellar objects may also yield interesting results. We also note that in the $V$ band, neglecting the effects of limb darkening (as we have done for these calculations) results in $R_p$ being overestimated by up to 50\% \citep[][Fig.~12]{SeagerMallen-Ornelas2003}. Therefore we deemed these candidates worthy of follow-up.

An inspection of Digitised Sky Survey images revealed that each of the candidates were affected by some level of blending in the APT photometry aperture. 

\begin{figure}
  \centering
  \includegraphics[width=63mm]{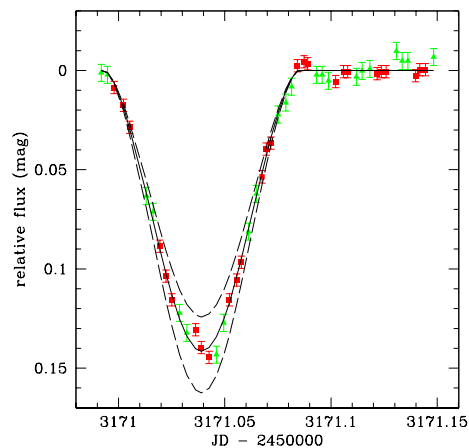}
  \caption{ The undiluted eclipse signal from transit candidate UNSW-TR-2, observed at high spatial resolution with the 40-inch telescope at SSO. We observed a single transit, and alternated between $V$ band (triangles) and $I$ band (squares). The solid line is the model lightcurve for an eclipsing system consisting of 2 K7V stars (with masses 0.58~\msun\ and radii 0.64~\msun) in a 2.12-day orbit, at an inclination of $\sim 83$\dg . The dashed lines show the result of varying the radius of the stars by $\pm 5$\% (keeping other parameters constant).}
  \label{fig:model}
\end{figure}

We obtained higher spatial resolution images using the 40-inch telescope at Siding Spring Observatory (SSO). We aimed to monitor each candidate continuously during a predicted transit, alternating between the $V$ and $I$ bands. To improve the time sampling we only read out one quarter of the full CCD, corresponding to a 10~arcmin square field (sampled by 1k$\times$1k, 0\farcs 6~pixels). Due to a run of bad weather, we only observed one of the candidates (UNSW-TR-2) during a transit. In this case the source of the transit signal is in fact the second bright star in the APT aperture ($\sim$0.8~mag fainter than the brightest one, and $\sim 10$\arcsec\ from it). The undiluted lightcurve (Fig.~\ref{fig:model}) shows a V-shaped eclipse with a depth of 0.14~mag, suggesting that the signal is from a grazing eclipsing binary system.

Simultaneously with the high-resolution imaging, we obtained spectra using the Double-Beam Spectrograph (DBS) on the 2.3~m telescope, also at SSO. We used the highest resolution grating in both arms. The approximate spectral ranges covered were 5700--6700~\AA\ in the red arm and 3900--4400~\AA\ in the blue, with resolutions of 1.1~\AA\ (45~\kmps) and 0.5~\AA\ (40~\kmps) respectively. We measured the radial velocities by cross-correlating the H$_\alpha$ absorption line in each spectrum with a spectrum of the RV standard star $\beta$ Vir (obtained with the same instrument).

We also obtained a single low-resolution (2.2~\AA) spectrum of each object covering the range 3420--5350~\AA. These were visually compared to templates from the UVILIB spectral library \citep{Pickles1998}, yielding a spectral type estimate for each candidate (Table~\ref{tab:followup}). 

Due to the limited observing time, we were unable to monitor all stars blended within each APT photometry aperture. Therefore we observed the brightest star in each aperture at several independent phases, and took single spectra of the fainter neighbouring stars. If the transits were in front of the fainter stars, they would be more likely to be of stellar origin (since the undiluted transit depth would have to be larger).

For three of the candidates (UNSW-TR-1, 3 and 4), the brightest star shows large RV variations in phase with the transit signal. The velocity semi-amplitude ($K$) of the candidates is shown in Table~\ref{tab:followup}. The companion masses were calculated using an estimate of the primary's mass from its spectral type, and assuming circular orbits (which is consistent with the radial velocity curve for two of these cases). Clearly the companions are low-mass stars, and not planets.

\subsubsection{A K7Ve binary system}\label{sssec:k7v}

For the fourth candidate (UNSW-TR-2), the radial velocity of the bright star dominating the light in the aperture showed no coherent variations when folded at the period measured from the transits. We only had one DBS spectrum of this star's close neighbour, which was revealed in the 40-inch images to be the source of the transit signal. This spectrum is well matched by a K7V template, with two important exceptions: (1) there is a strong H$_\alpha$ emission peak; and (2) the absorption lines and the emission line all have two components of approximately equal strength. Taking into account the orbital phase at the time the spectrum was taken, the maximum velocity separation between the components is $175 \pm 6$~\kmps.

We were able to obtain two more spectra of this star at lower resolution. Although the double line structure is not completely resolved in these spectra, they do confirm that the separation between the lines varies with orbital phase. 
We have not performed a rigorous fit to the lightcurves. However, a model consisting of two identical K7V stars, each of mass $0.58 \pm 0.05$~\msun\ and radius $0.64 \pm 0.01$~\rsun, appears to fit all the observations. The period is $2.11674 \pm 0.00002$~days (twice the original estimate), giving an orbital radius of $\sim 0.03$~AU. The orbital inclination is $83.4 \pm 0.1$~degrees.\footnote{The fitted radius and inclination are strongly correlated.}

This system may be a valuable addition to the small number of known double-lined eclipsing binary systems with late-type components, for which the physical parameters can be measured with precision (e.g. \citealt{TorresRibas2002,MaceroniMontalban2004}; Maceroni, C., private communication).
Such measurements provide important constraints for models of low-mass stars.

\subsection{Variable stars}

Visual inspection of the individual $V$-band lightcurves revealed the presence of 49 well-defined variable stars. Of these, 3 were previously known variable stars (HD~170451, BP~Ser and BI~Oph). We classified the remaining 46 following a simple procedure based on the period and the shape of the lightcurve. To minimise personal biases, we put stars into broad classes like short period pulsators or eclipsing binaries. For a few stars, we were able to determine the GCVS type unambiguously.

The following types of variables were found: 40 pulsating stars (33 short period, 5 RR~Lyrae-type, 2 long period semi-regular) and 9 eclipsing binaries. We found a number of bright short period pulsating stars having A--F spectral types\footnote{Spectral types were obtained from the SIMBAD online catalogue.}, which suggests that they are likely $\delta$~Scuti stars. For the majority, however, the lack of spectroscopic observations prevent more detailed classification.

The period determination was performed using Period98 of \citet{Sperl1998} and the phase dispersion minimisation method \citep{Stellingwerf1978}. In addition to the frequencies, we also calculated signal-to-noise  ratios in the Fourier spectra to get a measure of reliability of the  determined frequencies. Following \citet{Breger1993}, we adopted  $\sn > 4$ as the limit of significance.  The detected variable stars and their basic parameters (catalogue number, coordinates, period, type of variability) are listed in Table~\ref{tab:var}. 
Although in some cases, there is evidence for multiple periodicity, this is beyond the scope of the present paper.

\begin{table*}
\begin{minipage}{160mm}
  \caption{ Variable stars found in the NGC~6633 field. As this is a relatively crowded field, most objects are blends. As a consequence, (1) the actual variable star may be another star within $\sim 30$\as\ of the catalogue star we identify; (2) the coordinates shown represent the centroid of the blend; (3) the $V$ magnitudes measure the total flux in the APT photometry aperture; (4) the amplitudes may be diluted by blended stars, thus the values shown are only approximate lower limits. Coordinates were calculated from the WCS solution in the APT images, with a precision of $\sim 1$\as. The magnitude zero-point calibration is accurate to $\sim 0.1$ mag.}
  \label{tab:var}
  \centering
  \begin{tabular}{llcccccl}
    \hline
    Id & Catalogue Id         & R.A.       &  Dec.     & $V$  &  $P$     &   $A$       &   Type \\
       &                      & (J2000)    & (J2000)   &      &  (d)     & (mmag)   &                           \\ 
    \hline
V1     &                      & 18 22 10.2 &$+$6 23 20 & 12.9 &  0.09225 &        8 & short period pulsator     \\ 
V2     &                      & 18 22 08.3 &$+$6 42 42 & 13.2 &  0.78357 &      285 & W UMa                     \\ 
V3     &      GSC 00445-00528 & 18 23 11.7 &$+$6 18 53 & 10.1 &  0.13704 &       10 & short period pulsator     \\ 
V4     &                      & 18 23 06.9 &$+$6 42 13 & 12.3 &  0.14254 &       10 & short period pulsator     \\ 
V5     &      GSC 00445-00689 & 18 23 09.1 &$+$6 51 35 & 10.6 &  0.19818 &        9 & short period pulsator     \\ 
V6     &                      & 18 23 27.5 &$+$6 12 06 & 12.8 &  0.09452 &       42 & short period pulsator     \\ 
V7     &                      & 18 23 33.1 &$+$6 33 29 & 12.8 &  0.38673 &      242 & W UMa                     \\ 
V8     &                      & 18 23 47.9 &$+$7 28 05 & 12.4 &  1.02289 &      312 & Algol                     \\ 
V9     &      GSC 00445-01888 & 18 23 42.5 &$+$6 24 09 & 11.1 &  0.09855 &       10 & short period pulsator     \\ 
V10    &                      & 18 24 27.1 &$+$6 45 42 & 12.1 &  0.19167 &       13 & short period pulsator     \\ 
V11    &                      & 18 24 32.6 &$+$7 30 46 & 12.0 &  0.06285 &        5 & short period pulsator     \\ 
V12    &                      & 18 24 40.2 &$+$6 10 05 & 12.5 &  0.06559 &       42 & short period pulsator     \\ 
V13    &                      & 18 24 45.2 &$+$6 05 31 & 12.4 &  0.28535 &      153 & RR Lyrae                  \\ 
V14    &           NGC 6633 8 & 18 24 40.6 &$+$7 04 05 &  9.2 &  0.15436 &       38 & short period pulsator     \\ 
V15    &                      & 18 25 14.8 &$+$6 33 54 & 11.5 &  0.07567 &        6 & short period pulsator     \\ 
V16    &          NGC 6633 14 & 18 25 04.1 &$+$6 25 55 & 10.0 &  0.03658 &       14 & short period pulsator     \\ 
V17    &                      & 18 25 10.1 &$+$6 46 44 & 11.6 &          &          & eclipsing binary          \\ 
V18    &                      & 18 26 14.4 &$+$5 34 54 & 13.1 &  1.63336 &      185 & Algol                     \\ 
V19    &                      & 18 26 34.7 &$+$6 06 19 & 13.0 &  0.28913 &       36 & short period pulsator     \\ 
V20    &                      & 18 26 42.3 &$+$7 05 02 & 11.2 &  0.05681 &       20 & short period pulsator     \\ 
V21    &                      & 18 26 19.7 &$+$7 27 58 & 12.2 &  0.14068 &       10 & short period pulsator     \\ 
V22    &                      & 18 26 54.2 &$+$6 58 06 & 11.8 &  0.14862 &       45 & short period pulsator     \\ 
V23    &          NGC 6633 54 & 18 26 44.9 &$+$6 24 20 & 10.1 &  0.07472 &        4 & short period pulsator     \\ 
V24    &          NGC 6633 89 & 18 27 33.4 &$+$6 56 00 & 10.7 &  0.06734 &       10 & short period pulsator     \\ 
V25    &                      & 18 27 40.9 &$+$7 08 34 & 11.4 &  0.05363 &       18 & short period pulsator     \\ 
V26    &         NGC 6633 108 & 18 28 04.0 &$+$5 58 13 & 10.3 &  0.08375 &        5 & short period pulsator     \\ 
V27    &      GSC 00445-00634 & 18 27 53.3 &$+$6 08 51 & 11.2 &  0.34196 &       25 & RR Lyrae                  \\ 
V28    & Cl* NGC 6633 VKP 188 & 18 28 24.4 &$+$6 13 37 & 11.3 &  1.11068 &       17 & eclipsing binary          \\ 
V29    &                      & 18 28 57.3 &$+$6 21 02 & 13.3 &  1.81161 &      253 & eclipsing binary          \\ 
V30    &         NGC 6633 131 & 18 28 33.8 &$+$6 53 15 & 10.0 &  0.04723 &        8 & short period pulsator     \\ 
V31    &         NGC 6633 135 & 18 28 42.9 &$+$6 51 25 & 10.4 &  0.03634 &        6 & short period pulsator     \\ 
V32    &                      & 18 29 10.2 &$+$6 43 52 & 12.9 &  0.11138 &       24 & short period pulsator     \\ 
V33    &                      & 18 28 58.0 &$+$7 28 33 & 12.0 &  0.09568 &      188 & short period pulsator     \\ 
V34    &         NGC 6633 149 & 18 29 17.3 &$+$5 39 20 & 10.8 &  0.04683 &        6 & short period pulsator     \\ 
V35 & NGC 6633 147 (HD~170451)& 18 29 12.8 &$+$6 47 14 &  9.5 &  0.37532 &      355 & W UMa                     \\ 
V36    &            V* BP Ser & 18 30 13.4 &$+$6 16 50 & 12.5 &          &          &  LB                       \\ 
V37    &                      & 18 30 29.0 &$+$5 48 39 & 12.8 &  0.04203 &       12 & short period pulsator     \\ 
V38    &      GSC 00458-01442 & 18 30 41.9 &$+$6 47 50 & 10.9 &  0.31777 &        9 & RR Lyrae                  \\ 
V39    &      GSC 00458-00757 & 18 30 59.8 &$+$7 11 51 & 10.7 &  0.11008 &        4 & short period pulsator     \\ 
V40    &                      & 18 31 02.4 &$+$6 05 01 & 12.3 &  0.03941 &        7 & short period pulsator     \\ 
V41    &            V* BI Oph & 18 31 07.6 &$+$7 00 32 & 11.2 &  202.59  &          & Mira Cet                  \\ 
V42    &                      & 18 31 18.7 &$+$7 05 25 & 12.7 &  0.07343 &        9 & short period pulsator     \\ 
V43    &                      & 18 31 19.5 &$+$7 26 21 & 13.3 &  0.53548 &      186 & RR Lyrae                  \\ 
V44    &                      & 18 31 59.2 &$+$5 40 19 & 10.7 &  0.43263 &      122 & RR Lyrae                  \\ 
V45    &                      & 18 32 05.6 &$+$7 14 56 & 11.9 &  0.05936 &        9 & short period pulsator     \\ 
V46    &                      & 18 32 51.9 &$+$6 49 01 & 13.3 &  0.50552 &      156 & W UMa                     \\ 
V47    &                      & 18 33 17.3 &$+$5 58 31 & 12.2 &  0.21377 &      146 & pulsating                 \\ 
V48    &      GSC 00458-01108 & 18 33 16.4 &$+$6 29 08 &  9.1 &  0.15686 &        6 & short period pulsator     \\ 
V49    &                      & 18 34 03.5 &$+$6 36 03 & 12.6 &  0.13636 &       12 & short period pulsator     \\ 
    \hline
  \end{tabular}
\end{minipage}
\end{table*}


\section{Expected planet detection rate}\label{sec:drate}

\subsection{A rough estimate}\label{ssec:rough}

The number of detectable planets in our search can be approximated as 
\begin{equation} 
\ndet = \ndwarf \times \ppl \times  \ptr \times \pobs
\end{equation}
where \ndwarf\ is the number of dwarf stars in our fields for which a hot Jupiter would cause at least a 10~mmag transit, \ppl\ is the fraction of stars that actually host such a planet, \ptr\ is the geometric probability that the planet does transit (i.e. that it crosses our line of sight to the star), and \pobs\ is the probability that we observe multiple transits, with some minimum combined signal-to-noise ratio ($\snmin \approx 10$). 

A typical transiting hot Jupiter, with radius 1.2~\rjup\ \citep[e.g.][Fig.~3]{Sozzetti2004b}, will cause at least a 10-mmag transit in front of any star with radius $\rstar \le 1.2$~\rsun, i.e. of spectral type $\sim$~F8 or later.  For a magnitude-limited sample, $\sim 50$\% of stars are of type GKM \citep{Cox2000}. Thus we have $\ndwarf \approx 0.5 \times N_*$, where $N_*$ is the total number of stars monitored with at least 10-mmag precision (i.e. with $V \la 13$). For the NGC~6633 field, $N_* \approx 2000$.

\begin{figure}
  \includegraphics[angle=-90,width=84mm]{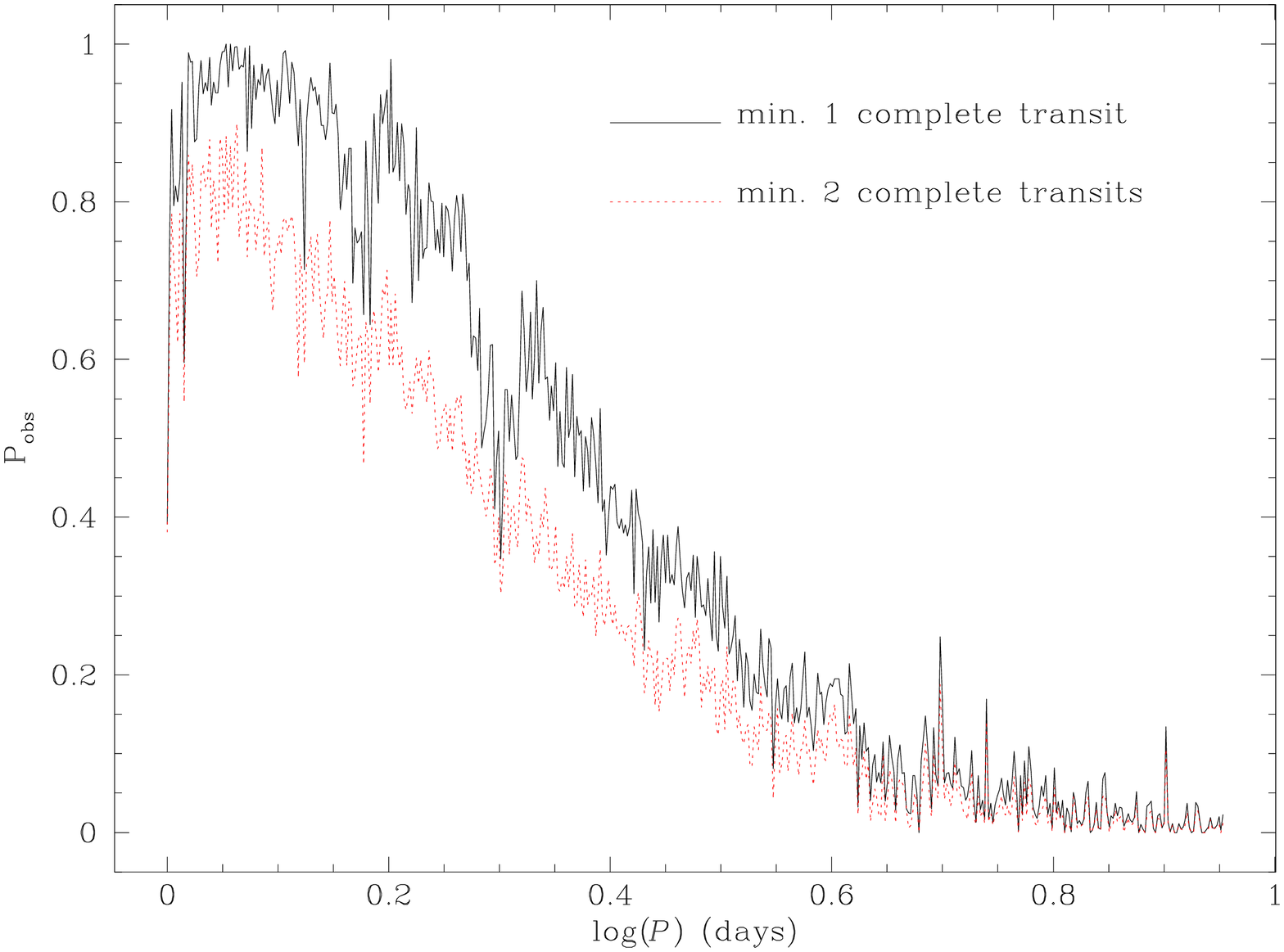}
  \caption{The probability of observing at least one (solid line) or two (dotted line) {\em complete} transits of a transiting planet in the lightcurves we have obtained for the NGC~6633 field. A minimum of 100 in-transit data points are required for a detection (so that 10-mmag precision yields a total signal-to-noise ratio of $\ge 10$ for a typical transiting hot Jupiter). A typical transit duration of 0.1~day is assumed. The one-transit probability also includes the requirement of at least one additional partial transit, so that the period can be estimated. The mean probability over this period range (1--9~days, sampled logarithmically) is 0.4 for one transit, and 0.3 for two transits.}
  \label{fig:pobs}
\end{figure}

The frequency of hot Jupiters is $\ppl \approx 1$\% \citep*[e.g.][]{GaudiSeagerMallen-Ornelas2004}, and their transit probability is $\ptr \approx \rstar/a \sim 0.1$ ($a$ is the orbital radius). Therefore, observing a rich Galactic field like NGC~6633, we have $\ndet \approx \pobs$. For our complete NGC~6633 lightcurves, $\pobs \approx 0.4$ (minimum one complete and one partial transit, $1 < P < 9$~days, see Fig.~\ref{fig:pobs}), giving an estimate of 0.4 planets for this dataset, or $\sim 0.2$ planets per year.

In this simple estimate, we have made some gross simplifications, including
\begin{enumerate}
\item neglecting the dependence of \ppl, \ptr\ and \pobs\ on the orbital period, and other properties of the star and planet;
\item assuming a very simple (and perhaps optimistic) dependence of the detection probability on the transit depth, and neglecting the distribution of transit depths among target systems;
\item neglecting the 3-dimensional distribution of stars in the Galactic disk;
\item neglecting interstellar extinction;
\item neglecting the effect of blends in a crowded field;
\item ignoring the possible effects of stellar companions on detection probability (only $\sim 1/3$ of late-type dwarfs are single stars, \citealt{DuquennoyMayor1991}).
\end{enumerate}
Therefore this is only an order-of-magnitude figure, but will provide a useful comparison to the results of our more detailed analysis.

 We note that \citet{Horne2003} estimates a detection rate of 3 planets per month for our project, using simple scaling laws. However, by his calculations, all the transit surveys combined should be detecting 191 planets each month, which is clearly an overestimate. The main goal of these calculations was perhaps not an estimate of the absolute detection rate, but rather a comparison of the various projects. If we assume the six planets found by the transit method to date were detected from a single year of observations by all the projects, and scale Horne's estimate to match this rate, then his estimate for our search becomes 0.09 planets per year. For a transit search with parameters comparable to ours (though a shorter observing run), \citet{Brown2003} estimates a rate of $\sim 0.7$ detected planet per $10^4$ stars monitored (to a magnitude limit of $R = 12$). Our estimate above translates to $\sim 2$ planets per $10^4$ stars. Brown's analysis includes a number of effects, such as that of the planet period and radius distributions, which reduce the detection rate compared to our simple estimate.

\subsection{Monte Carlo simulation}

In order to obtain a more realistic estimate, we have modelled our observations using a Monte Carlo simulation. This allows us to use a realistic photometric noise model, and to take various observational effects into account (in particular blending, and the set of epochs at which the lightcurves are sampled).

We begin by simulating a population of stars in a 1 square degree field of view. We use the luminosity function for the Solar neighbourhood, $\Phi_0(M_V)$ (as tabulated in \citealt{Cox2000}, for $M_V = -7, -6, ... 18$), allowing for an exponential drop-off with height above the Galactic plane (with scale heights, $H(M_V)$, from \citealt{Allen1973}). The number of stars of absolute magnitude $M_V$ per cubic parsec, at a distance $d$ and Galactic latitude $b$, is thus
\begin{equation}
\Phi(M_V, d, b) = \Phi_0(M_V) e^{-d|\sin b|/H(M_V)}  
\end{equation}

We divide the volume sampled by a 1 square degree field into a series of shells of thickness 20~pc. We fill each shell with the appropriate number of stars in each $M_V$ bin. To calculate the apparent ($V$) magnitude of each star, we assume an extinction of 1.9 magnitudes per kiloparsec in the Galactic plane, dropping off exponentially with distance from the plane, with a scale height of 140~pc \citep{Allen1973}. Only stars with $5 < V < 19$ are included in the model, as stars brighter than this can be avoided during field selection, while fainter stars are not resolved in APT images.

Depending on $M_V$, an appropriate fraction of stars (tabulated in \citealt{Cox2000}) are simulated as binary systems. For simplicity, a luminosity ratio is randomly selected from a uniform distribution in the range (0, 1)\footnote{The observed distribution of binary mass ratios peaks near $q=0.3$ and drops off considerably at higher values \citep{DuquennoyMayor1991}. Therefore on average, our simplification overestimates the dilution of the primary's light by the companion, leading to a slightly lower estimate of the probability of detecting a planet around the primary.}. The absolute magnitude of the primary is adjusted for this (keeping the total luminosity of the system equal to the original $M_V$), and the secondary is ignored.

Apart from binary companions, a  3-pixel radius photometry aperture centred on a star in an APT image typically includes light from several other stars. We randomly assign pixel coordinates to the simulated stars, within a section of the APT CCD corresponding to one square degree. For each star, we then calculate what fraction of the total starlight in the aperture comes from that star, and take this into account when calculating the observed transit depth due to a planet. Stars for which this fraction is less than 5\% are excluded from further analysis, as any planetary transit signal from these stars would be diluted below our detection limit.

Around each remaining star, we place a planet with a radius of $\rplanet=1.2$~\rjup. We randomly select a period in the range 1--9 days, using the distribution described by \citet{GaudiSeagerMallen-Ornelas2004}: uniform in $\log P$ in each of two bins, 1--3~d (``very hot Jupiters'', VHJ) and 3--9~d (``hot Jupiters'', HJ), with the frequency of VHJ relative to HJ being 15\%.

Tables in \citet{Lang1992} give the mass (\mstar) and radius (\rstar) of the star based on its absolute magnitude, assuming it is on the main sequence. For the planet, we assume a circular orbit, with radius $a = (\mbox{1~AU})(\mstar / \msun)^{1/3}(P/(\mbox{1~yr}))^{2/3}$. We then randomly select an orbital inclination ($i$) leading to a transit (i.e. select $\cos i$ from a uniform distribution in the range 0--\ptr, where $\ptr \approx \rstar / a$ is the probability of such an inclination for a randomly oriented orbit).

Neglecting limb-darkening, grazing transits, and ingress/egress times, we approximate the signal as a rectangular transit of depth $\trdep = (\rplanet/\rstar)^2$, and duration
\begin{equation} 
\Delta t = \frac{P}{\pi a} \sqrt{ (\rstar+\rplanet)^2 - (a \cos i)^2 }
\end{equation}
We assume that we will reliably detect planets with $\trdep > \trdepmin = 10$~mmag, provided multiple transits are observed and a minimum total signal-to-noise ratio is reached ($\snmin=10$). Considering that the lightcurves we have obtained to date frequently contain systematic trends of order 10~mmag, this value of \trdepmin\ is somewhat optimistic. For our present analysis of the NGC~6633 lightcurves, a value of 20~mmag would be more realistic. Well-sampled transits of that depth (superimposed on the systematics) are easily detectable, at least by visual inspection. However, we are confident that our implementation of the \citet{KovacsBakosNoyes2004} trend filtering algorithm, combined with transit detection software, will allow us to reliably detect transits as shallow as 10~mmag (Sec.~\ref{ssec:cansel}). Thus we will continue to use this optimistic value for \trdepmin, while noting that it is an important parameter. Its effect on the planet catch is discussed below.

We estimate the photometric precision per measurement using a noise model which includes Poisson noise due to star and sky flux, and a limiting precision of 1.5~mmag for the bright stars (the solid curve in Fig.~\ref{fig:rms}). Combined with \trdep, this tells us the number of in-transit measurements (\nmin) that need to be made in order to achieve a total signal-to-noise ratio \snmin. Given a simulated planet and a set of observation epochs, we can calculate the probability (\pobs) that a minimum of 2 complete transits will be observed, including at least \nmin\ in-transit points. We do this by testing 1000 transit epochs.

To estimate the total planet catch for the assumed observing parameters (Galactic latitude and observation epochs), we add up the values of $\ptr \times \pobs$ for each simulated planet, and scale the result to the appropriate field size. We also multiply by the fraction of stars that do host short-period planets, which is $\sim 1$\% \citep[e.g.][]{GaudiSeagerMallen-Ornelas2004}.

\subsection{Validating the simulation}

In order to ensure that this rather complicated simulation yields reasonable results, we performed a number of tests. Firstly, we generated a set of fake APT images using the distribution of stars from the simulation (for $b \approx 8$\dg). The simulated images appear very similar to a real image at the same Galactic latitude (with the exception of some patches of increased extinction in the real image). Processing these fake images through the photometry pipeline, we also confirmed that the star counts as a function of limiting magnitude closely match those obtained from a real image.

We then modified the simulation to reproduce the simplifying assumptions made by \citet{Horne2003}. The result was within a factor of 2 of Horne's estimate for our project. Finally, we confirmed that the result scales with planet, star, and observational parameters (\rplanet, $a$, $M_V$, \rstar, sky brightness, survey duration and \snmin) as predicted by Horne's equation~3.

\subsection{Results for the NGC~6633 field}

We have simulated observations at the Galactic latitude of our NGC~6633 field ($b = 8$\dg), using the epochs of the best images we have obtained for this field  in 2002--2004 (2486 images, from 61 nights over 3 seasons). The expected planet catch for a single 6-square-degree APT field is $\sim 0.1$ planet ($\trdepmin=10$~mmag, 2 complete transits). If we accept one complete observed transit as a detection (plus at least one partial transit so that some estimate of the period can be made), the estimate becomes $\sim 0.2$ planet, which agrees well with the rough estimate derived in Section~\ref{ssec:rough}.

\begin{figure}
  \includegraphics[width=84mm]{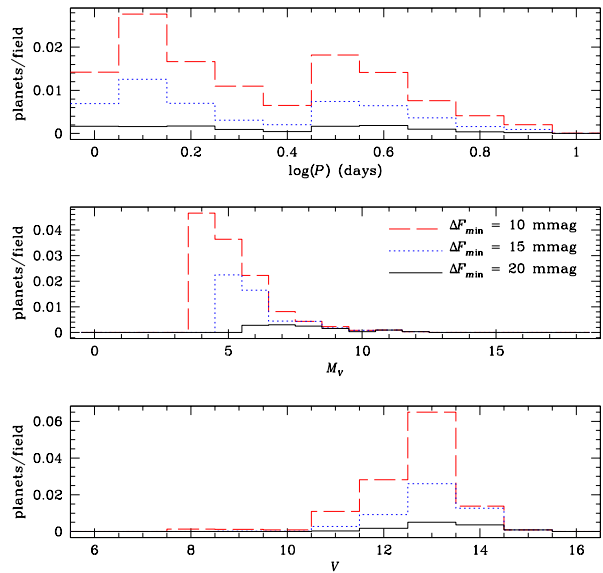}
  \caption{Distribution of period ({\em top panel}), host star absolute magnitude ({\em middle panel}) and apparent magnitude ({\em bottom panel}) for detected extrasolar planets in the Monte Carlo simulation. The histograms were generated by weighting each simulated planet by the probability of detecting it ($\ptr \times \pobs$). The simulation parameters were set to match those of our observations of the NGC~6633 field in 2002--2004. In each panel we show the distribution for three different values of the minimum detectable transit depth.}
  \label{fig:simold}
\end{figure}

As expected, we are most sensitive to planets with the shortest periods (Fig.~\ref{fig:simold}). However, since we have simulated planets with periods longer than 3~d to be almost 10 times more frequent than those with $P<3$~d (according to the observed distribution), the distribution of detected planets also has a peak at 3--5~d. Most detected planet hosts have an apparent magnitude $V \approx 13$, while almost none are fainter than $V\approx14.5$.

The luminosity distribution of detected host stars peaks at $M_V \approx 4$ (for $\trdepmin=10$~mmag), as also predicted by \citet*{PepperGouldDepoy2003}. Amongst the stars that are small enough for a 1.2~\rjup\ planet to cause a 10~mmag transit, those with $M_V \approx 4$ are the brightest and therefore the most numerous in our magnitude-limited sample. As a result, \trdepmin\ is perhaps the most important parameter in determining the total planet catch.\footnote{The largest detectable host-star radius is $\trdepmin^{-1/2}\rplanet$. Varying the assumed planet size (\rplanet) is equivalent to varying $\trdepmin^{-1/2}$.}
In the case of the NGC~6633 dataset, about 90\% of the systems detected with $\trdepmin=10$~mmag have transits shallower than 20~mmag (Table~\ref{tab:trdep}). This highlights the importance of removing systematic trends from our lightcurves so that shallow signals can be detected.

\begin{table}
  \caption{The total number of detected planets per APT field (minimum two complete observed transits) as a function of the minimum detectable transit depth (\trdepmin, given in the first row in mmag). We compare two simulations: our NGC~6633 observations (2002--2004, $V$ band, Sec.~\ref{ssec:fields}), and higher-cadence observations at a Galactic latitude of 25\dg\ in $I$ band over two months (Sec.~\ref{ssec:disc}).}
  \label{tab:trdep}
  \centering
  \begin{tabular}{lcccc}
    \hline
    Strategy     &   10 &   15 &   20 &   30  \\ %
    \hline
    NGC~6633 obs.& 0.12 & 0.05 & 0.01 & 0.004 \\ %
    New strategy & 0.20 & 0.12 & 0.08 & 0.050 \\ %
    \hline
  \end{tabular}
\end{table}

Repeating the simulation for one of the other 7 fields (for each of which we have $\sim 1000$ useful images, over 2 seasons) gives 0.03 planets per field ($\trdepmin=10$~mmag, 2 complete transits). The reduced estimate compared to the NGC~6633 field is due to the smaller number of images, fewer nights with long continuous coverage (mostly due to bad weather), and sparser time sampling. Thus our most optimistic estimate of the planet catch in the data we collected over 2002--2004 is $\sim 0.3$ planets, or $\sim 0.6$ if we settle for one complete transit.

\subsection{Discussion}\label{ssec:disc}

The above disappointing estimate suggests that our initial strategy of simultaneously monitoring four fields over multiple seasons is far from optimal. We selected this strategy believing it would maximise the number of stars we monitor, while obtaining a minimum time-sampling required to detect a transit (one image every $\sim14$ minutes). The signal-to-noise ratio (\sn) of a detection would be increased by folding the lightcurve at the appropriate period. However, gaps in the observations due to weather and technical problems have reduced the probability of detecting multiple, complete transits (Fig.~\ref{fig:pobs}). To date, the systematic trends present in many lightcurves have prevented the use of detection software to make shallow transits detectable by phase-folding. We need to sample the lightcurves more frequently so that the required total \sn\ can be accumulated over a small number of observed transits.

\begin{figure}
  \includegraphics[width=84mm]{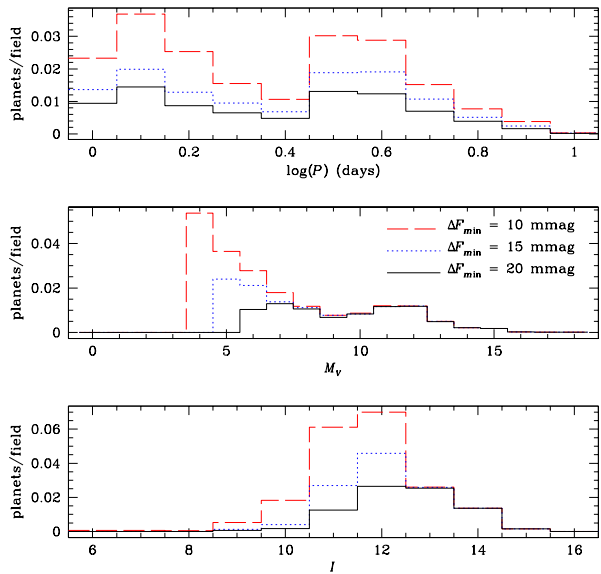}
  \caption{Distribution of the properties of detected systems for a model of our optimal observing strategy. Compare to Fig.~\ref{fig:simold}, representing our initial strategy. We are confident that with future improvements in our trend filtering and detection software, the $\trdepmin=10$~mmag case will become a realistic one.}
  \label{fig:simnew}
\end{figure}

We have used the Monte Carlo simulations to study the effects of various observational parameters on the planet catch, and thus select a more efficient search strategy. One important change was the choice of filter. Other transit-search teams have discussed the merits of observing at long wavelengths, in particular in $I$ band \citep[e.g.][]{vonBraun2005}. In terms of maximising the detection rate, there are two main advantages. First, the greater CCD sensitivity in $I$ band allows shorter exposures (60~s, compared to 150~s in $V$) and thus higher-cadence observations. Second, observing in $I$ band significantly increases the number of well-measured red dwarf stars in our sample (which have deeper transits) relative to their larger, bluer counterparts. This makes the total detection rate less sensitive to the minimum detectable transit depth (see Table~\ref{tab:trdep} and Fig.~\ref{fig:simnew}).

The simulations indicate that the detection rate is essentially constant from the Galactic plane out to a latitude of $|b| \approx 25$\dg, and is only reduced by $\sim 20$\% even at $|b| \approx 65$\dg. This is due to the reduced blending in less crowded fields, as well as the near-isotropic distribution of the well-measured red stars (most of which are within $\sim100$~pc). Although we have not calculated the frequency of ``false positive'' detections due to eclipsing binaries blended with a third star along the line of sight, this will be significantly lower in the higher latitude fields. Furthermore, the photometric precision (and absence of systematics) we have assumed in the simulation are easier to achieve in less crowded fields. For these reasons we conclude that fields in the range $15\dg \la |b| \la 45\dg$ are optimal. 

Future fields will be at southern declinations, so that they are observable at low airmass for a full 8 hours every night.\footnote{From SSO, we have a minimum of 8 hours of darkness per night for 8 months each year.} We will select pairs of adjacent fields, obtaining an image of each field every 4 minutes. 

We have simulated this new strategy, taking into account the different stellar magnitudes, lower interstellar extinction and brighter sky in $I$ band. We can achieve a detection rate of 0.2 planet per field ($\trdepmin=10$~mmag, 2 complete transits) in only 20 nights of observations (taken over up to 2 months, allowing for gaps due to bad weather and full Moon).

Observing 10 fields per year (one pair for every 2 months, allowing for telescope downtime), will lead to an overall detection rate of 2 planets per year, a significant improvement over the estimated 0.2--0.3 per year for our 2002--2004 observations. If further efforts do not allow us to reliably detect transits shallower than 20~mmag, the estimate for the new strategy is halved, while for the past observations it is reduced by a factor of 10 (Table~\ref{tab:trdep}).


\section{Conclusions}\label{sec:conlc}

We have described in detail the methods we have used to achieve millimagnitude-precision relative photometry using the Automated Patrol Telescope. For bright stars, the precision was limited by intra-pixel variations in CCD sensitivity (IPSV), coupled with undersampling of the images. This was largely overcome by a new observing technique, which optimally alters the effective PSF in order to average out the effects of IPSV. We have demonstrated that using this technique, a customised aperture photometry package and an automated processing pipeline, we can obtain the large number of high-quality lightcurves required to detect transiting extrasolar planets.

During 2002--2004, we observed 8 fields near the Galactic plane. Our analysis of data from the field centred on the open cluster NGC~6633 yielded 4 transit candidates. Follow-up observations revealed each of these to be eclipsing binary systems. We also found 49 variable stars in this field. Three of the other fields also contained binary stars with eclipses similar to those of a transiting planet.

By simulating the population of stars we observe, the properties of the telescope, and our observing strategy, we have studied the sensitivity of our search to short-period planets. Assuming the frequency of planets around stars targeted by radial velocity surveys is representative of the Galactic field stars we observe, we have estimated our expected detection rate. Considering all the good-quality lightcurves for the 8 fields we have observed to date, we expect to find 0.3--0.6 planets.

Our new strategy, whereby we obtain high-cadence observations in $I$ band, will yield an estimated 2 planets per year, a 10-fold increase in detection rate over previous observations. We have been using this strategy since November 2004. The new camera for the APT, being built in 2005, will have at least 3 times the field of view of the current one, higher sensitivity, and smaller pixels. This will further increase the detection rate by a factor of a few.


\section*{Acknowledgments}
The authors wish to thank Sun Microsystems for their generous donation of a workstation. We are grateful to our referee, Frederic Pont, whose comments significantly improved the Section~\ref{sec:drate} of this paper. We also thank Suzanne Aigrain for shared code and useful discussions; Chris Blake and Anna Frebel for using some of their own observing time to obtain spectra for us; Brad Carter, Carla Maceroni, Tom Marsh and Pierre Maxted for their valuable comments; and the Mount Stromlo and Siding Spring Observatories TAC for allocations on the 2.3-m and 40-inch telescopes. AD wishes to thank L{\'a}szlo Kiss for useful comments and suggestions. MGH and JLC are supported by Australian Postgraduate Awards. AD is supported by the International Postgraduate Research Scholarship programme of the Australian Department of Education, Science and Training. The APT camera upgrade is funded by the Australian Research Council. The NASA ADS Abstract Service was used to access data and references. This research has made use of the SIMBAD and VizieR databases, operated at CDS, Strasbourg, France.


\bibliographystyle{mn2e}
\bibliography{Hidas2005}

\bsp

\label{lastpage}

\end{document}